\documentclass[aps,prl,floatfix,twocolumn,reprint,amsmath,amssymb,superscriptaddress,UTF8]{revtex4-2}
\usepackage{amsfonts}
\usepackage{mathrsfs}
\usepackage{amsmath}% needed for subequations
\usepackage{color}
\usepackage{natbib}
\usepackage{graphicx}
\usepackage{bm}% bold maths
\usepackage{amssymb}
\usepackage{xspace}
\usepackage{epstopdf}
\usepackage{dcolumn}% Align table columns on decimal point
\usepackage{longtable}
\usepackage{array}
\usepackage{makecell}
\usepackage{tabulary}
\usepackage{multirow}
\newcolumntype{C}[1]{>{\centering\arraybackslash}p{#1}}
\newcolumntype{L}[1]{>{\flushleft\arraybackslash}p{#1}}

\def\ie{{\it i.e.},\ }

\usepackage{multirow}
\usepackage{epsfig}
\usepackage[normalem]{ulem}
\usepackage{amsmath}
\usepackage{braket}
\usepackage{bbold}
\usepackage{natbib}

\usepackage[colorlinks=true, letterpaper=true, pdfstartview=FitV, linkcolor=blue, citecolor=blue, urlcolor=blue]{hyperref}

\makeatletter

\newcommand{\Rmnum}[1]{\expandafter\@slowromancap\romannumeral #1@}
\makeatother
\begin{document}

\title{Symmetry Classification of Altermagnetism and Emergence of Type-IV Magnetism in Two Dimensions}

\author{Mu Tian}
\thanks{These authors contributed equally to this work.}
\affiliation{Key Lab of advanced optoelectronic quantum architecture and measurement (MOE), Beijing Key Lab of Nanophotonics $\&$ Ultrafine Optoelectronic Systems, and School of Physics, Beijing Institute of Technology, Beijing 100081, China}
\affiliation{International Center for Quantum Materials, Beijing Institute of Technology, Zhuhai, 519000, China}

\author{Chaoxi Cui}
\thanks{These authors contributed equally to this work.}
\affiliation{Key Lab of advanced optoelectronic quantum architecture and measurement (MOE), Beijing Key Lab of Nanophotonics $\&$ Ultrafine Optoelectronic Systems, and School of Physics, Beijing Institute of Technology, Beijing 100081, China}
\affiliation{International Center for Quantum Materials, Beijing Institute of Technology, Zhuhai, 519000, China}

\author{Zeying Zhang}
\affiliation{College of Mathematics and Physics, Beijing University of Chemical Technology, Beijing 100029, China}

\author{Jingyi Duan}
\affiliation{Key Lab of advanced optoelectronic quantum architecture and measurement (MOE), Beijing Key Lab of Nanophotonics $\&$ Ultrafine Optoelectronic Systems, and School of Physics, Beijing Institute of Technology, Beijing 100081, China}
\affiliation{International Center for Quantum Materials, Beijing Institute of Technology, Zhuhai, 519000, China}

\author{Wanxiang Feng}
\affiliation{Key Lab of advanced optoelectronic quantum architecture and measurement (MOE), Beijing Key Lab of Nanophotonics $\&$ Ultrafine Optoelectronic Systems, and School of Physics, Beijing Institute of Technology, Beijing 100081, China}
\affiliation{International Center for Quantum Materials, Beijing Institute of Technology, Zhuhai, 519000, China}

\author{Run-Wu Zhang}
\email{zhangrunwu@bit.edu.cn}
\affiliation{Key Lab of advanced optoelectronic quantum architecture and measurement (MOE), Beijing Key Lab of Nanophotonics $\&$ Ultrafine Optoelectronic Systems, and School of Physics, Beijing Institute of Technology, Beijing 100081, China}
\affiliation{International Center for Quantum Materials, Beijing Institute of Technology, Zhuhai, 519000, China}

\date{\today}
\begin{abstract}
Two-dimensional (2D) magnetism,  particularly 2D altermagnetism (AM), has attracted considerable interest due to its exceptional physical properties and broad application potential. However, the classification of AM undergoes a fundamental paradigm shift when transitioning from three-dimensional (3D) to 2D symmetry-enforced fully compensated collinear magnetism$-$a shift that has remained largely overlooked. Here, by extending unconventional magnetism to 2D collinear systems, we identify the symmetry conditions  and electronic band characteristics of a distinct magnetic phase: type-IV magnetism. This new class lies beyond the established descriptions of ferromagnetism, conventional antiferromagnetism, and AM. Type-IV magnetism supports the successive emergence of both nonrelativistic spin-degenerate and relativistic spin-splitting phenomena, belonging strictly to neither conventional antiferromagnetism nor standard AM. We further establish a universal symmetry classification framework for 2D type-IV magnets via a mapping from the collinear spin layer group to the magnetic layer group. Monolayer MgCr$_2$O$_3$ and monolayer BaMn$_2$Ch$_3$ (Ch=Se, Te) are showcased as representative materials, exhibiting gate-tunable reversible spin textures and the quantum electric Hall effect, respectively. Our work underscores the rich functional prospects of type-IV magnets, offering a new route toward spin manipulation and anomalous transport that promises innovative designs for high-performance spintronic devices.

\end{abstract}
\maketitle

%%%%%%%%%%%%%%%%%%%%%%%%%%%%%%%%%%%%%%%
\textit{\textcolor{blue}{Introduction.-}}With the advance of diverse spin-dependent magnetic transport phenomena in antiferromagnetic materials~\cite{lebrunTunableLongdistanceSpin2018,chengSpinPumpingSpinTransfer2014,zhangSpinHallEffects2014,wangAntiferromagnonicSpinTransport2014,frangouEnhancedSpinPumping2016,zeleznySpinPolarizedCurrentNoncollinear2017,zeleznySpinTransportSpin2018,hortensiusCoherentSpinwaveTransport2021,rezendeDiffusiveMagnonicSpin2016,houSpinTransportAntiferromagnetic2019}, antiferromagnetic spintronics has driven considerable strides in spintronics~\cite{zuticSpintronicsFundamentalsApplications2004a,10.1063/5.0184580,baltzAntiferromagneticSpintronics2018a,jungwirthAntiferromagneticSpintronics2016,smejkalTopologicalAntiferromagneticSpintronics2018}. This progress has reignited interest in classifying antiferromagnetism (AFM). Of particular themes are unconventional antiferromagnets~\cite{liuDifferentFacetsUnconventional2025a,chenAnomalousHallEffect2014a,smejkalCrystalTimereversalSymmetry2020a,tsaiElectricalManipulationTopological2020,wadleyElectricalSwitchingAntiferromagnet2016,smejkalGiantTunnelingMagnetoresistance2022,nakatsujiLargeAnomalousHall2015,zhuObservationPlaidlikeSpin2024,gonzalezbetancourtSpontaneousAnomalousHall2023,takagiSpontaneousTopologicalHall2023,fengTopologicalMagnetoopticalEffects2020,maMultifunctionalAntiferromagneticMaterials2021,yuanGiantMomentumdependentSpin2020,guoQuantumAnomalousHall2023a}, enjoying antiferromagnetic configurations yet ferromagnetic properties, offering many fascinating performances such as high storage capacity, high packing densities, high-frequency operation and ultrafast dynamics. Altermagnetism (AM) represents a distinctive class of symmetry-enforced fully-compensated collinear magnetic orders. These are characterized by nonrelativistic spin splitting in momentum space~\cite{smejkalConventionalFerromagnetismAntiferromagnetism2022a}. Currently, AM emerges as a third fundamental magnetic phase alongside ferromagnetism (FM) and conventional AFM. This completes the classification of nonrelativistic collinear magnetism within the framework of spin-space group theory~
\cite{chenEnumerationRepresentationTheory2024b,jiangEnumerationSpinSpaceGroups2024a,liuSpinGroupSymmetryMagnetic2022a,xiaoSpinSpaceGroups2024a}. Compared to FM and conventional AFM, AM incorporates nearly all advantages of both, thereby unlocking phenomena and functionalities beyond the reach of conventional magnetic systems~\cite{adamantopoulosSpinOrbitalMagnetism2024,baduraObservationAnomalousNernst2024,chakrabortyChargeEntanglementCriticality2024,devarajInterplayAltermagnetismPressure2024,hanNonvolatileAnomalousNernst2025,hodtInterfaceinducedMagnetizationAltermagnets2024,huCatalogCPairedSpinMomentum2025,krempaskyAltermagneticLiftingKramers2024b,liCreationManipulationHigherorder2024,linObservationGiantSpin2024,liuTwistedMagneticVan2024,mcclartyLandauTheoryAltermagnetism2024,osumiObservationGiantBand2024,ouassouDcJosephsonEffect2023,parfenovPushingAltermagnetUltimate2025,reimersDirectObservationAltermagnetic2024,shaoNeelSpinCurrents2023,smejkalEmergingResearchLandscape2022,yuanDegeneracyRemovalSpin2023,yuanUncoveringSpinorbitCouplingindependent2023,zhouCrystalThermalTransport2024a,zengDescriptionTwodimensionalAltermagnetism2024,sodequistTwodimensionalAltermagnetsHigh2024,sheoranSpontaneousAnomalousHall2025b}. 

Two-dimensional (2D) AM retains the advantages above while offering exceptional structural flexibility for miniaturized quantum devices and exhibiting exotic physical properties not found in conventional magnets. However, transitioning from three-dimensional (3D) to 2D cases constitutes a fundamental paradigm shift in AM classification$-$one that has received insufficient attention. While 3D AM is rigorously defined, dimensional reduction fundamentally alters the specific spin group symmetry influence, disrupting the nonrelativistic description of AM in lower dimensions. Take 2D collinear magnets defined in the $xy$ plane as an example, the absence of the $k_z$ direction in their Brillouin zone (BZ) ensures that the spin group symmetries $\{C_2||2_z\}$ and $\{C_2||m_z\}$ ($C_2$ denotes a 180$^\circ$ rotation around an axis perpendicular to the spins) enforce spin degeneracy at every $k$ point, which is a feature absent in 3D case. Although spin space groups remain indispensable for describing nonrelativistic phenomena (especially in AM), their dimensional fragility, manifested through fundamentally distinct band degeneracies induced by spin-group operations from the 3D to 2D cases, renders them inadequate for classifying nonrelativistic phenomenology in reduced dimensions. Therefore, a new collinear spin layer group (cSLG) theory is urgently required for rigorous classification of 2D collinear AM.

In this Letter, we utilize the cSLGs theory to elucidate 2D collinear AM, classifying the 158 cSLGs into two categories, corresponding to 92 type-III cSLGs (\textit{i.e.}, 2D AM) and 66 type-IV cSLGs (\textit{i.e.}, type-IV magnetism). 
As an emerging unconventional 2D collinear magnetic phase, type-IV magnetism supports nonrelativistic spin-degenerate phenomena, thereby transcending the description of the conventional AM paradigm. In addition, we provide a complete materials design principle: a mapping framework between cSLGs and magnetic layer groups (MLGs), which is equally important to identify candidate materials. Through first-principles calculations, we predict two candidate materials: monolayer MgCr$_2$O$_4$ and monolayer BaMn$_2$Ch$_3$ (Ch=Se, Te). For monolayer MgCr$_2$O$_4$, the clean band structure and spin-layer coupling (SLC) enable gate-field-controlled switching of perfect spin textures. In monolayer BaMn$_2$Ch$_3$ (Ch=Se, Te), a novel quantum electric Hall effect~\cite{cuiElectricHallEffect2024} emerges within the type-IV magnetism, where the Hall conductivity exhibits quantized plateaus at 0 and $\pm$1 (in units of e$^2$/h) under the gate-field modulation. Our work not only presents a new paradigm for 2D unconventional type-IV magnets but also introduces a highly efficient and convenient method for its generation and control, due to the great tunability of the electric gate-field, thus opening a new frontier in spintronics, and stimulating broad interest in both theoretical exploration and practical applications.

\begin{figure}[h]
	\begin{center}  
		\includegraphics[width=0.5\textwidth]{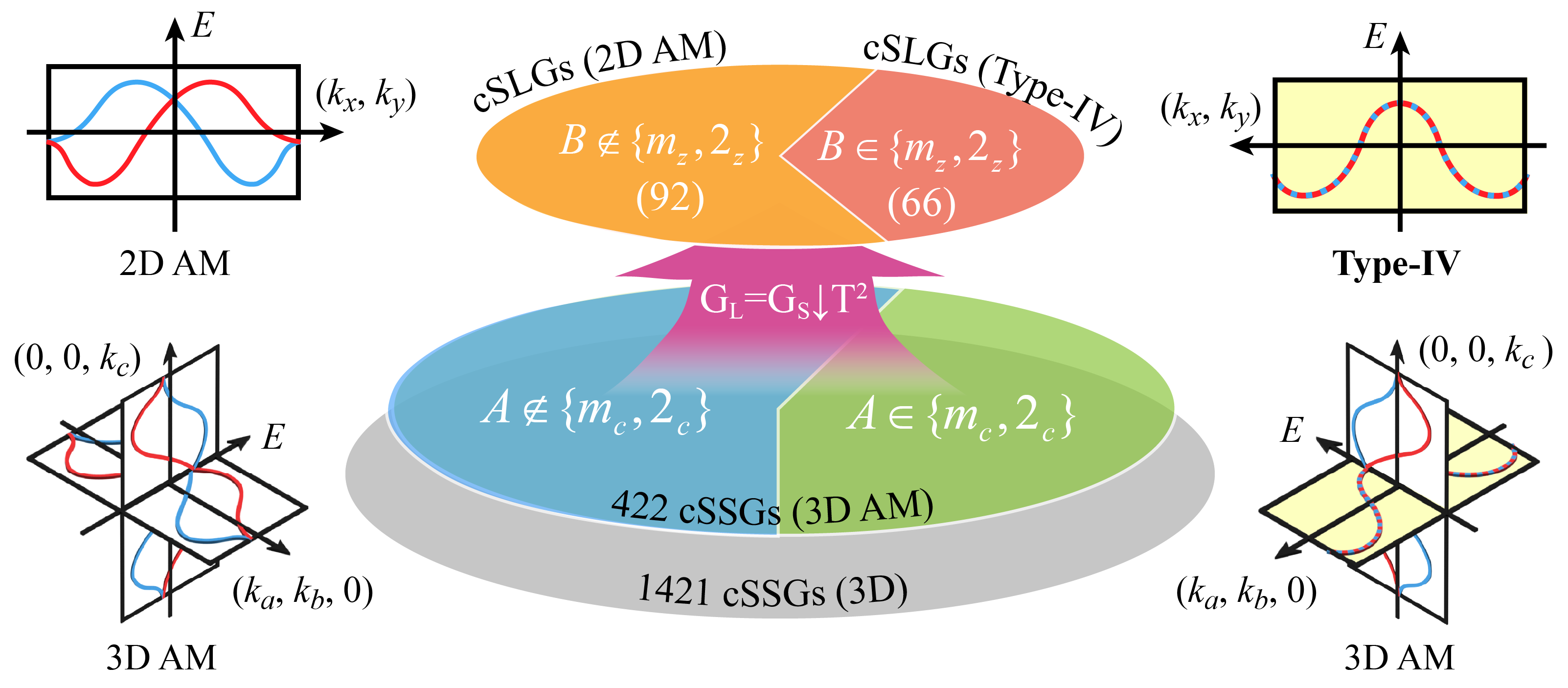}
		\caption{New classification of AM based on collinear spin layer groups (cSLGs). For the 3D case, 422 out of the 1421 collinear spin space groups (cSSGs) belong to AM. However, transitioning from 3D to 2D cases represents a fundamental paradigm shift in AM classification. For the 2D case, 92 cSLGs belong to 2D AM, and 66 cSLGs belong to type-IV magnetism.} \label{fig:1}  
	\end{center}
\end{figure}

\textit{\textcolor{blue}{New classifications of 2D AM.-}}
Advances in understanding collinear spin space groups (cSSGs) enable comprehensive classification of the 1,421 nontrivial cSSGs for 3D collinear magnets (denoted as $G_{NS}^l$) into four distinct categories based on their group element characteristics~
\cite{chenEnumerationRepresentationTheory2024b,jiangEnumerationSpinSpaceGroups2024a,liuSpinGroupSymmetryMagnetic2022a,xiaoSpinSpaceGroups2024a}. 
Among these, Type-III SSGs (\textit{viz}., cSSGs for 3D AM ) are described by $G_{AM}=\{E||G_S\}+\{C_2||AG_S\},A\in\{C_n,\mathcal{P}C_n\}$, where $G_S$ is a normal subgroup of the space group $G_S+AG_{S}$ and $A$ denotes the symmetry operation connecting the sublattices with opposite spins. Besides, $C_n$ and $\mathcal{P}C_n$ ($n=2, 4$) stand for $n$-fold symmorphic or nonsymmorphic space rotations and rotoinversions, respectively. In 1,421 cSSGs, there are 422 inequivalent cSSGs for AM in total to describe the recently emerged altermagnets with nonrelativistic spin splitting.

While cSSGs remain essential for describing AM, their dimensional sensitivity produces fundamentally different spin-group-induced band degeneracies between 3D and 2D systems~\cite{SM}. Regarding the 3D AM, the space operation $A$ that connects different spin sublattices is crucially important. If $A=\mathcal{O}_c\in\{M_c,C_{2,c}\}$, where $c$ is the axis perpendicular to $a-b$ plane, a spin-degenerate $k_a-k_b$ nodal plane is enforced at $k_c=0$ plane\cite{smejkalConventionalFerromagnetismAntiferromagnetism2022a}. It is noted that the spin-momentum locking is influenced not only by operations in $G_{AM}$ but also operations in the spin-only group of collinear magnets, which is denoted as: $G_{SO}^l=C_\infty+\overline{C}_2C_\infty$. Group $C_\infty=\{U(\phi),\phi \in [0,2\pi)\}$ contains all full spin rotations along a common axis, which guarantees spin as a good quantum number. Symmetry operation $\overline{C}_2$ is accompanied by a time reversal ($\mathcal{T}$) in real space that flips the sign of crystal momentum $\mathbf{k}$. Therefore, for the $\mathbf{k}$-dependent bands $E(s,\mathbf{k})$ of all collinear magnets, $E(s,\mathbf{k})=E(s,-\mathbf{k})$ is enforced by symmetry $\{\overline{C}_2||\mathcal{T}\}$. For energy spectrum distributed on $k_a-k_b$ plane, we have:
 \begin{equation}
     \begin{aligned}
         &\{C_2||M_c\}E(s,\mathbf{k})=E(-s,\mathbf{k})\\
         &\{\overline{C}_2||\mathcal{T}\}\{C_2||C_{2,c}\}E(s,\mathbf{k})=E(-s,\mathbf{k}).
     \end{aligned}
     \label{equ:symm}
 \end{equation} 
As shown in the green part of cSSGs for 3D AM in Fig.~\ref{fig:1}, Eq.(\ref{equ:symm}) enforces the spin-degenerate nodal plane.

This critical limitation precludes their application to reduced-dimension nonrelativistic classification, leaving 2D collinear AM without a formal cSLG theoretical framework. To solve this problem, we derived the cSLGs from the cSSGs, thereby reclassifying the 2D AM. In principle, the symmetry operations of all cSLGs are inherited from 3D cSSGs. Therefore, all possible cSLGs can be obtained by constraining cSSGs with periodicity in only two directions.  By treating cSLGs related by origin shifts or coordinate transformations as equivalent, we have systematically identified a total of 448 inequivalent cSLGs. The detailed algorithm can be found in Ref.~\cite{zeying}. 
After processing all cSSGs of 3D AM and removing equivalents, we identify 158 inequivalent cSLGs with the $z$-axis as the non-periodic direction, which is denoted as $G_{AM}^{2D}=\{E||G_L\}+\{C_2||BG_L\},B\in\{C_n,\mathcal{P}C_n\}$ ($G_L$ stands for the layer group). Among these, 66 cSLGs satisfying $B\in \{M_z,C_{2,z}\}$  enforce nonrelativistic spin degeneracy throughout the entire BZ. This exceptional behavior arises from dimensional reduction. For 2D collinear magnets with $\mathbf{k}=(k_x,k_y)$, Eq.(\ref{equ:symm}) remains valid when $c=z$, ensuring full BZ spin degeneracy. We designate the 66 cSLGs exhibiting this property (red region, Fig.~\ref{fig:1}) as cSLGs of type-IV magnetism, highlighting their connection to yet distinction from AM (orange region, Fig.~\ref{fig:1}).

\textit{\textcolor{blue}{Guide to 2D type-IV magnets design.-}}
Recent advances in collinear magnetism based on Laue group theory offer a framework for classifying magnetic systems and exploring the anomalous Hall effect~\cite{Bai2025Anomalous}. However, this spin-point-group approach remains retrospective and lacks a systematic protocol for predicting new materials$-$especially in two dimensions. In contrast, this work employs the cSLG formalism to establish a predictive materials design framework, enabling symmetry-based identification of realizable 2D type-IV magnets.

Moreover, unlike 3D magnets, the Mermin-Wagner theorem~\cite{merminAbsenceFerromagnetismAntiferromagnetism1966,halperinHohenbergMerminWagner2019a} prohibits spontaneous symmetry breaking in 2D systems, which necessitating spin-orbit coupling (SOC) to stabilize long-range magnetic order~\cite{merminAbsenceFerromagnetismAntiferromagnetism1966,halperinHohenbergMerminWagner2019a,sodequistTwodimensionalAltermagnetsHigh2024}. This constraint underscores the importance of establishing a one-to-one mapping between cSLGs and MLGs, thereby providing a symmetry-guided framework for designing 2D type-IV magnets.

Realizing type-IV magnets require magnetic moments aligned along a fixed axis and exclusive consideration of symmetry operations acting on these moments, enabling correspondences between all 448 cSLGs and MLGs. Here, we align moments along the $z$-axis to construct a comprehensive type-IV magnetism symmetry mapping table, as this orientation optimally preserves crystalline symmetry, maintaining $C_{4z}$ symmetry in square lattices and $C_{3z}$ symmetry in hexagonal lattices, while permitting fully compensated antiferromagnetic ordering along the $z$-direction across all type-IV magnetism systems.

\begin{figure}[h]
	\begin{center}  
		\includegraphics[width=0.5\textwidth]{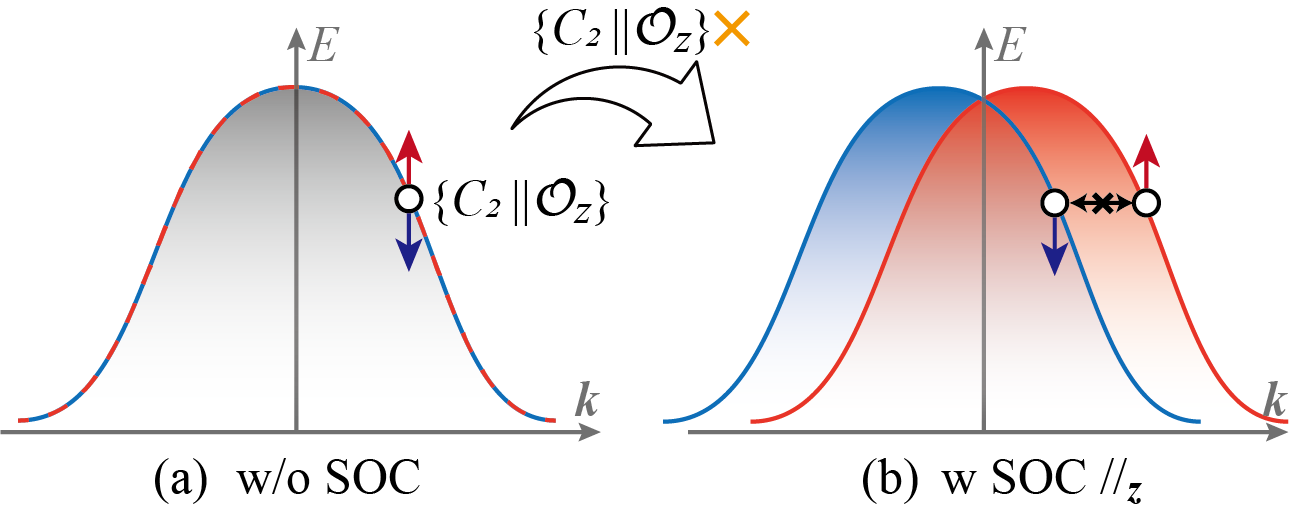}
		\caption{Band structures of a type-IV magnet in (a) nonrelativistic and (b) relativistic cases. Breaking the spin-degeneracy-enforcing symmetry $\{C_2||\mathcal{O}_z\}$ leads to generally spin-split bands in the relativistic regime, induced by SOC.} \label{fig:2}  
	\end{center}
\end{figure}

With the $z$-axis magnetic moment orientation specified, operations within nontrivial cSLGs can be directly mapped to MLGs operations, as detailed in Table S1~\cite{SM}. Utilizing this correspondence table, we can determine the associated MLGs for any of the 66 cSLGs for type-IV magnetism; these one-to-one mappings are provided in Table S2~\cite{SM}. By leveraging the established $(\mathrm{cSLG, MLG})$ pairs, we systematically investigate the symmetries of type-IV magnetism in both nonrelativistic and relativistic regimes. This approach establishes a practical framework for screening materials to identify candidates with desired properties.
Among the MLGs, due to the absence of $\{C_2||\mathcal{O}_z\}$ compared to cSLGs, there is no symmetry maintaining the degeneracy of spin. Generally, the relativistic band of type-IV magnetism is consequently spin-splitting, as illustrated in Fig. \ref{fig:2}.

Alongside developing design principles, identifying candidate materials is equally essential. The unique symmetry constraints of type-IV magnetism grant them rich physical properties. Guided by theoretical analysis, we took monolayer MgCr$_2$O$_4$ and monolayer BaMn$_2$Te$_3$ as examples to demonstrate that these two candidate materials can achieve gate-field controllable spin textures and quantum electric Hall effect.

\begin{figure}[h]
	\begin{center}  
		\includegraphics[width=0.5\textwidth]{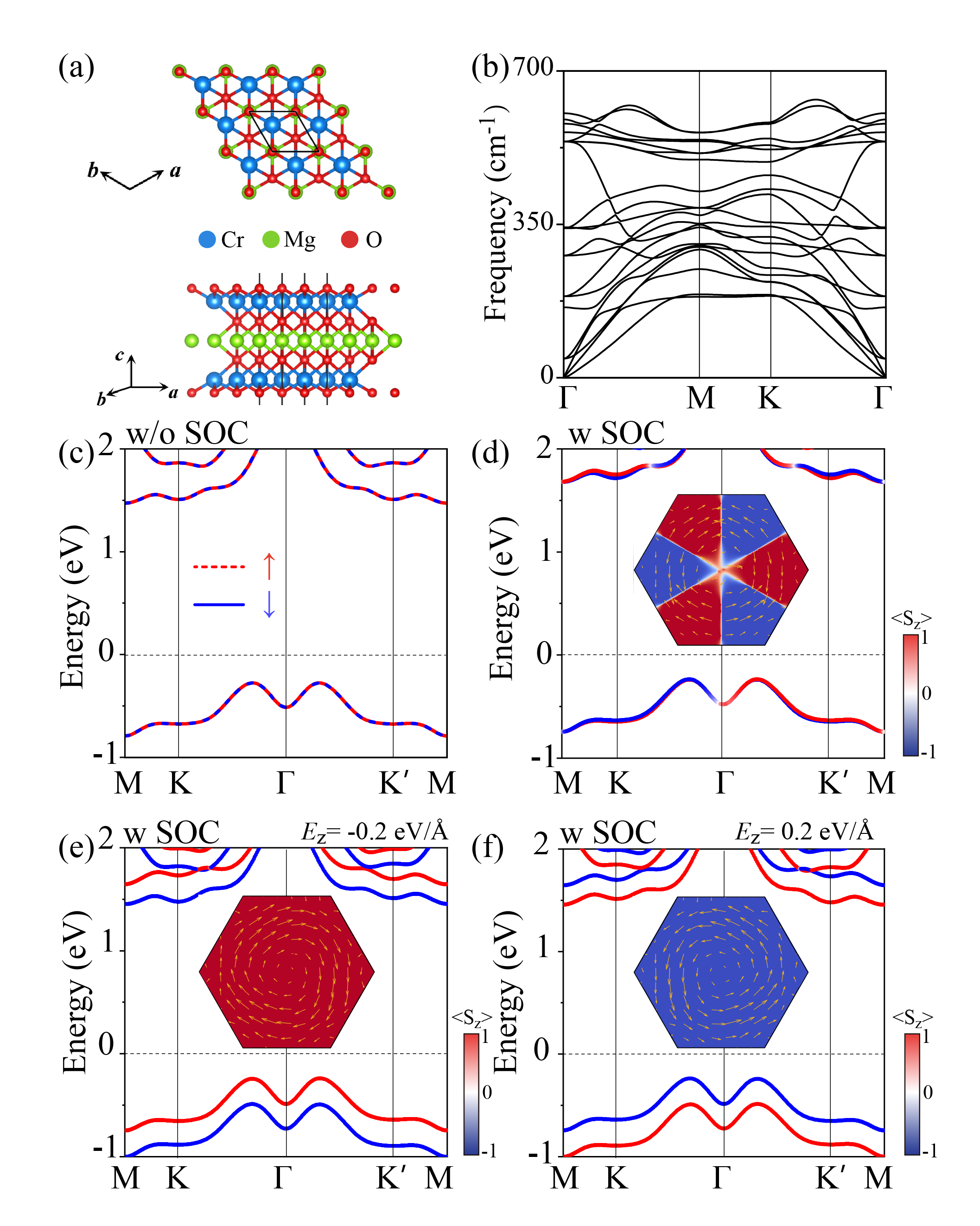}
		\caption{(a) Relaxed geometry of monolayer MgCr$_2$O$_4$. (b) Phonon spectrum of the monolayer MgCr$_2$O$_4$. (c) Nonrelativistic band structure of monolayer MgCr$_2$O$_4$ in N$\acute{\mathrm{e}}$el configuration. The red dashed line and blue solid line denote the spin-up channel and the spin-down channel, respectively. (d) $s_z$-resolved relativistic band structure of monolayer MgCr$_2$O$_4$ with the N$\acute{\mathrm{e}}$el vector perpendicular to the plane. The inset illustrates the spin texture of the valence band, where the blue and red regions denote the sign of $\langle {s}_z\rangle$ and the yellow arrows denote the in-plane spin texture. (e) $s_z$-resolved relativistic band structure of monolayer MgCr$_2$O$_4$ under a gate field of $E_z$=0.2 eV. The inset illustrates the spin texture of the valence band. (f) $s_z$-resolved relativistic band structure of monolayer MgCr$_2$O$_4$ under a gate field of $E_z$=-0.2 eV. The inset illustrates the spin texture of the highest valence band.} \label{fig:3}  
	\end{center}
\end{figure}

\textit{\textcolor{blue}{Candidate 1: MgCr$_2$O$_4$-}}
The geometric structure of monolayer MgCr$_2$O$_4$ is depicted in Fig.~\ref{fig:3}(a), revealing a seven-layer atomic sequence of O-Cr-O-Mg-O-Cr-O. One Mg atom lies in the middle plane, sandwiched between the two CrO$_2$ layers.
Such 2D material can be obtained via mechanical exfoliation from bulk MgCr$_2$O$_4$ (mp-2225288), with an exfoliation energy of about 0.3 J/m$^2$, which is comparable to those of graphene~\cite{hanSurfaceEnergiesAdhesion2019,jungRigorousMethodCalculating2018a,wangMeasurementCleavageEnergy2015} and MoS$_2$~\cite{rascheDeterminationCleavageEnergy2022a}.
The crystal lattice belongs to the layer group $P\overline{6}m2$ (No. 78), with an optimized lattice constant of $a=b=2.99$ \AA.
Furthermore, monolayer MgCr$_2$O$_4$ demonstrates dynamic stability [see Fig.~\ref{fig:3}(b)] and retains thermal stability up to 300 K~\cite{SM}, implying that MgCr$_2$O$_4$ is robust enough to form a freestanding 2D material.
By comparing the energies of potential magnetic configurations, we confirm that the ground state of the monolayer MgCr$_2$O$_4$ exhibits a N$\acute{\mathrm{e}}$el-type type-IV magnetism~\cite{SM}. 
The magnetic moments are mainly on the Cr sites with a magnitude of $\sim 3 \mu_B$, with the N\'eel vector aligned along the $z$ axis, resulting in a maximum value of the magnetic anisotropy energy of 0.356 meV/Cr~\cite{SM}.

In the absence of SOC, the monolayer MgCr$_2$O$_4$ belongs to the cSLG No. $187.1.2.2.L.1$ (see details of Sec.II~\cite{SM}). Its symmetry operation $\{C_2||M_z\}$ connects opposite spin sublattices and, as dictated by Eq.(\ref{equ:symm}), enforce $E(s,\mathbf{k})=E(s,-\mathbf{k})$, thereby leading to the spin degeneracy across the entire 2D BZ [Fig.~\ref{fig:3}(c)]. This phenomenon is significantly different from the definition of AM.
When SOC is included, the system transitions to MLG No. $78.5.514$, the breaking of symmetry $\{C_2||M_z\}$ lifts the spin degeneracy of band structure, creating a clear signature of spin-splitting [Fig.~\ref{fig:3} (d)]. Since $C_\infty$ symmetry of spin is broken under the constraint of MLG No. $78.5.514$, spin ceases to be a good quantum number, generating finite in-plane components $\langle {s}_x\rangle$ and $\langle {s}_y\rangle$ alongside the out-of-plane $\langle {s}_z\rangle$, as shown in the inset in Fig.~\ref{fig:3} (d).

Achieving all-electrical control of spin textures in type-IV magnets would mark a fundamental advance toward next-generation devices that unify memory and computing capabilities~\cite{guoElectricallySwitchablePersistent2023,khomitskyElectricfieldInducedSpin2009,krempaskyOperandoImagingAllElectric2018}. Through symmetry analysis, we demonstrate type-IV magnet MgCr$_2$O$_4$ as a potential candidate hosting SLC~\cite{SM}. Owing to SLC, one can easily predict that the appearance of reversible spin textures driven under opposite perpendicular gate fields ($\pm E_{z}$) is predictable. 
Applying a positive gate field (\textit{e.g.}, $E_z$ = +0.2~eV/\text{\AA}), the calculated band structure shown in Fig.~\ref{fig:3}
(e) reveals a significant ($\sim 0.25 $ eV) spin splitting of the two highest valence bands within $+\langle {s}_z\rangle$ and $-\langle {s}_z\rangle$ components. The inset of Fig.~\ref{fig:3} (e) shows more specifically a 2D view of the spin textures of the highest valence band with $+\langle {s}_z\rangle$ component, illustrating the clockwise helical in-plane pattern, reflecting the dominance of Rashba effect over Dresselhaus effect in monolayer MgCr$_2$O$_4$. When an opposite gate field is applied (\ie $E_z$ = -0.2~eV/\text{\AA}), the highest valence band features $-\langle {s}_z\rangle$ component, showcasing the counterclockwise helical in-plane pattern [Fig.~\ref{fig:3} (f)].

\begin{figure}[h]
	\begin{center}  
		\includegraphics[width=0.5\textwidth]{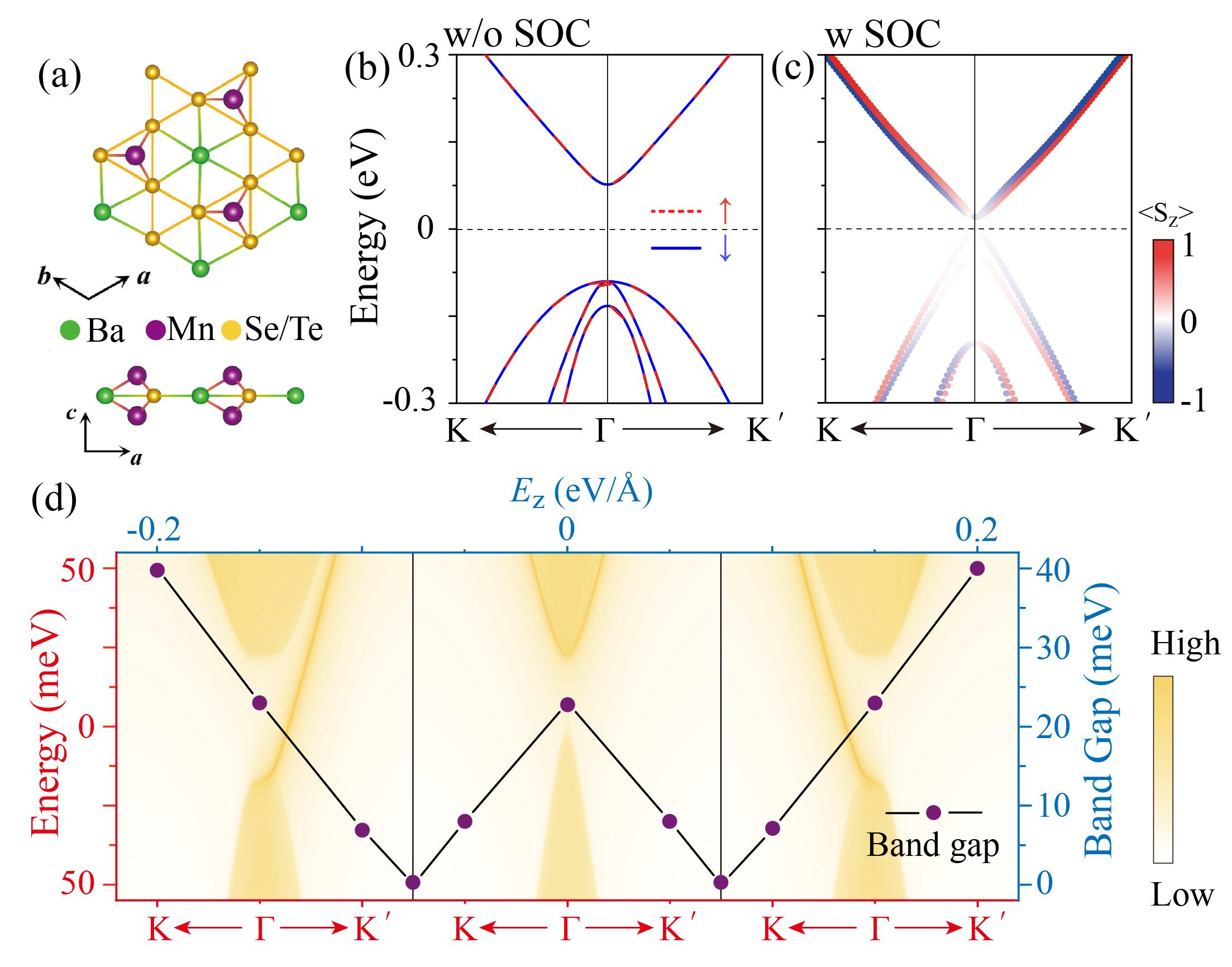}
		\caption{(a) Top and side views of monolayer BaMn$_2$Ch$_3$ (Ch = Se, Te). (b) Nonrelativistic band structure of monolayer BaMn$_2$Te$_3$. The red dashed line and blue line denote the spin-up channel and the spin-down channel, respectively. (c) Corresponding $s_z$-resolved relativistic band structure of monolayer BaMn$_2$Te$_3$. (d) Band gap of relativistic band structure of monolayer BaMn$_2$Te$_3$ (right axis) as a function of gate field $E_z$ (up axis). The edge states for $E_z=\pm0.2$~eV/\text{\AA} are plotted as background (left and down axes). } \label{fig:4}  
	\end{center}
\end{figure}        

\textit{\textcolor{blue}{Candidate 2: BaMn$_2$Te$_3$.-}}
We further demonstrate monolayer BaMn$_2$Ch$_3$ (Ch=Se, Te) materials as type-IV magnets [Fig.~\ref{fig:4}(a)], with BaMn$_2$Te$_3$ emerging as the most representative system. Detailed analysis of BaMn$_2$Te$_3$ is presented in the main text, while corresponding results for monolayer BaMn$_2$Se$_3$ are provided in the Supplemental Material~\cite{SM}. Monolayer BaMn$_2$Te$_3$ is structured in a hexagonal lattice with lattice constant $a=b=7.67$ \AA. The magnetic configuration of monolayer BaMn$_2$Te$_3$ in the ground state with N\'eel vector oriented perpendicular to the plane~\cite{SM}. The Mn atoms account for the majority of the magnetic moment at approximately 4.4 $\mu_B$ per atom.

As another candidate, monolayer BaMn$_2$Te$_3$ falls into the class of 2D type-IV magnetism. Regarding the nonrelativistic case, it belongs to the cSLG No. $187.1.2.2.L.1$, displaying the spin degeneracy band structure due to the $\{C_2||M_z\}$ symmetry [Fig.~\ref{fig:4}(b)]. 
When SOC is considered, the band structure of monolayer BaMn$_2$Te$_3$ undergoes significant changes, from the original spin degeneracy to obvious spin splitting, showing a semiconductor with a direct gap from the original 167 meV to 22 meV at $\varGamma$ point, as shown in Fig.~\ref{fig:4}(c).

The band structure of monolayer BaMn$_2$Te$_3$ can also be manipulated by the gate field. When a perpendicular electric field $E_z$ is applied, the system's band gap becomes tunable and exhibits dependence on the field strength (See Fig.~\ref{fig:4}(d)). 
The band gap closes when $E_z $ is set with a proper strength $ \pm E_c$ with $E_c=0.075$~eV/\text{\AA}. When the strength of the gate field $|E_z|$ increases, the band gap reopens and the system transitions into a topological insulator phase, as evidenced by the chiral edge states plotted as background of Fig.~\ref{fig:4}(d). 
Notably, applying an electric field of $E_z=\pm0.2$~eV/\text{\AA} enhances the system's band gap to $\sim 40 $ meV, significantly exceeding the characteristic thermal energy at room temperature ($\sim 26 $ meV).
This electric generation and  control of Hall conductivity may serve as a quantized version of the electric Hall effect, which is named as Quantum Electric Hall Effect~\cite{cuiElectricHallEffect2024}. Furthermore, Fig.~\ref{fig:4}(d) clearly demonstrates that the polarity of the applied electric field can selectively control chiral edge states with opposite handedness.

\textit{\textcolor{blue}{Conclusion.-}}
We extend the concept of unconventional magnetism to 2D collinear magnets, thereby identifying a distinct magnetic phase: type-IV magnetism. Going beyond the definition of AM, this emergent magnetic phase enables the successive emergence of nonrelativistic spin-degenerate and relativistic spin-splitting phenomena in type-IV magnets. This behavior is comparable to that of a magnetic analogue of MoS$_2$ and it falls under neither conventional AFM nor AM. Also, we establish a universal symmetry rule to categorize 2D type-IV magnets by mapping the relation of cSLGs and MLGs. Proposing concrete materials, monolayer MgCr$_2$O$_4$ and monolayer BaMn$_2$Ch$_3$ (Ch=Se, Te), as two candidates for realizing 2D type-IV magnets with unique SLC and intriguing effects within the gate-field control of spin and the quantum electric Hall effect~\cite{cuiElectricHallEffect2024}. These findings not only provide new perspectives for exploring novel type-IV magnetism but also offer a promising platform for designing next-generation spintronic devices.

\bibliography{4thmag}

\end{document}